\documentclass{article}
\usepackage{amssymb, amsmath, amsfonts, graphicx, color, bm, float}
\usepackage{natbib}
\usepackage{hyperref}
\DeclareMathOperator{\E}{E}

\DeclareMathOperator{\Cov}{Cov}

\DeclareMathOperator{\argmax}{arg max}
\DeclareMathOperator{\logit}{logit}

\newcommand{\yobs}{y_{\text{obs}}}
\newcommand{\sobs}{s_{\text{obs}}}
\newcommand{\ABC}{\text{ABC}}

\newcommand{\onlyChapter}[1]{}
\newcommand{\onlyarXiv}[1]{#1}

\title{Summary Statistics in Approximate Bayesian Computation}
\author{Dennis Prangle\footnote{Newcastle University.  Email \href{mailto:dennis.prangle@gmail.com}{\nolinkurl{dennis.prangle@newcastle.ac.uk}}}}
\date{}

\begin{document}

\maketitle

\begin{abstract}
This document is due to appear as a chapter of the forthcoming Handbook of Approximate Bayesian Computation (ABC) edited by S. Sisson, Y. Fan, and M. Beaumont.

Since the earliest work on ABC, it has been recognised that using summary statistics is essential to produce useful inference results.
This is because ABC suffers from a curse of dimensionality effect, whereby using high dimensional inputs causes large approximation errors in the output.
It is therefore crucial to find low dimensional summaries which are informative about the parameter inference or model choice task at hand.
This chapter reviews the methods which have been proposed to select such summaries, extending the previous review paper of \cite{Blum:2013} with recent developments.
Related theoretical results on the ABC curse of dimensionality and sufficiency are also discussed.
\end{abstract}

\section{Introduction}

To deal with high dimensional data, ABC algorithms typically reduce it to lower dimensional \emph{summary statistics} and accept when simulated summaries $S(y)$ are close to the observed summaries $S(\yobs)$.
This has been an essential part of ABC methodology since the first publications in the population genetics literature.
Overviewing this work \cite{Beaumont:2002} wrote
``A crucial limitation of the\ldots method is that only a small number of summary statistics can usually be handled.
Otherwise, either acceptance rates become prohibitively low or the tolerance\ldots must be increased, which can distort the approximation'',
and related the problem to the general issue of the \emph{curse of dimensionality}:
many statistical tasks are substantially more difficult in high dimensional settings.
In ABC the dimension in question is the number of summary statistics used. 

To illustrate the issue, consider an ABC rejection sampling algorithm.
As more summary statistics are used there are more opportunities for random discrepancies between $S(y)$ and $S(\yobs)$.
To achieve a reasonable number of acceptances it is necessary to use a large threshold and accept many poor matches.
Therefore, as noted in the quote above, using too many summary statistics distorts the approximation of the posterior.
On the other hand, if too few summary statistics are used some fine details of the data can be lost.
This allows parameter values to be accepted which are unlikely to reproduce these details.
Again a poor posterior approximation is often obtained.

As a result of the considerations above, a good choice of ABC summary statistics must strike a balance between low dimension and informativeness.
Many methods have been proposed aiming to select such summary statistics,
and the main aim of this chapter is to review these.
There is some overlap between this material and the previous review paper of \cite{Blum:2013}.
This chapter adds coverage of recent developments, particularly on auxiliary likelihood methods and ABC model choice.
However less detail is provided on each method here, due to the larger number now available.
The chapter focuses on summary statistic selection methods which can be used with standard ABC algorithms.
Summary statistic methods for more specialised recent algorithms \onlyChapter{(see chapters 8, 11 and 21 of this book for example} \onlyarXiv{\citep[e.g.][]{Ratmann:2013, Barthelme:2014, Drovandi:2015}} are discussed only briefly.
Secondary aims of the chapter are to collate relevant theoretical results and discuss issues which are common to many summary statistic selection methods.

An overview of the chapter is as follows.
Section \ref{sec:theory} is a review of theoretical results motivating the use of summary statistics in ABC.
Section \ref{sec:strategies} describes three strategies for summary statistic selection and introduces some general terminology.
Sections \ref{sec:subsetselection}-\ref{sec:auxlik} describe particular methods from each strategy in turn.
Up to this point the chapter concentrates on ABC for parameter inference.
Section \ref{sec:modelchoice} instead considers summary statistics for ABC model choice, covering both theory and methods.
Section \ref{sec:discussion} concludes by summarising empirical comparisons between methods, discussing which method to use in practice and looking at prospects for future developments.

\section{Theory} \label{sec:theory}

This section describes why methods for selecting summary statistics are necessary in more theoretical detail.
Section \ref{sec:cod} discusses the curse of dimensionality, showing that low dimensional informative summaries are needed. 
The concept of \emph{sufficient statistics}, reviewed in Section \ref{sec:sufficiency},
would appear to provide an ideal choice of summaries.
However it is shown that low dimensional sufficient statistics are typically unavailable.
Hence methods for selecting low dimensional insufficient summaries are required.

Note that from here to Section \ref{sec:auxlik} the subject is ABC for parameter inference, understood to mean inference of \emph{continuous} parameters.
Theoretical results on ABC for model choice will be discussed in Section \ref{sec:modelchoice}.
These are also relevant for inference of discrete parameters.

\subsection{The curse of dimensionality} \label{sec:cod}

A formal approach to the curse of dimensionality is to consider how the error in an ABC approximation is related to the number of simulated datasets produced, $n$.
It can be shown that, at least asymptotically, the rate at which the error decays becomes worse as the dimension of the data increases.
For example \cite{Barber:2015} consider mean squared error of a Monte Carlo estimate produced by ABC rejection sampling.
Under optimal ABC tuning and some regularity conditions this is shown to be $O_p(n^{-4/(q+4)})$, where $q$ denotes $\dim S(y)$.
This is an asymptotic result in a regime where $n$ is large and the ABC bandwidth $h$ is close to zero.
Several authors \citep{Blum:2010, Fearnhead:2012, Biau:2015} consider different definitions of error and different ABC algorithms, and prove qualitatively similar results.
That is, similar asymptotic expressions for error are found with slightly different terms in the exponent of $n$.
While these asymptotic results may not exactly capture behaviour for larger $h$, they strongly suggest that high dimensional summaries typically give poor results.

Note that it is sometimes possible to avoid the curse of dimensionality for models whose likelihood factorises, such as state space models for time series data.
This can be done by performing ABC in stages for each factor.
This allows summary statistics to be chosen for each stage, rather than requiring high dimensional summaries of the entire model.
\cite{Jasra:2015} reviews this approach for time series data, and a related new method is in chapter 21 of this book.

\subsection{Sufficiency} \label{sec:sufficiency}

Two common definitions of sufficiency of a statistic $s=S(y)$ under a model parameterised by $\theta$ are as follows.
See \cite{Cox:1979} for full details of this and all other aspects of sufficiency covered in this section.
The classical definition is that the conditional density $\pi(y|s,\theta)$ is invariant to $\theta$.
Alternatively, the statistic is said to be \emph{Bayes sufficient} for $\theta$ if $\theta|s$ and $\theta|y$ have the same distribution for any prior distribution and almost all $y$.
The two definitions are equivalent for finite dimensional $\theta$.
Bayes sufficiency is a natural definition of sufficiency to use for ABC, as it shows that in an ideal ABC algorithm with sufficient $S$ and $h \to 0$, the ABC target distribution equals the correct posterior.
It can also be used to consider sufficiency for a subset of the parameters which is useful later when ABC model choice is considered.

For independent identically distributed data,
the Pitman-Koopman-Darmois theorem states that under appropriate assumptions only models in the exponential family possess a sufficient statistic with dimension equal to the dimension of $\theta$.
For other models the dimension of any sufficient statistic increases with the sample size. 
Exponential family models generally have tractable likelihoods so that ABC is not required.
This result strongly suggests that for other models low dimensional sufficient statistics do not exist.

Despite this result there are several ways in which notions of sufficiency can be useful in ABC.
Firstly, stronger sufficiency results are possible for ABC model choice, which are outlined in Section \ref{sec:modelchoice}.
Secondly, notions of approximate and asymptotic sufficiency are used to motivate some methods described later (i.e.~those of \citealp{Joyce:2008} and \citealp{Martin:2014}).

\section{Strategies for summary statistic selection} \label{sec:strategies}

This chapter splits summary statistic selection methods into three strategies: \emph{subset selection}, \emph{projection} and \emph{auxiliary likelihood}.
This section gives an overview of each and introduces some useful general terminology.
The categories are based on those used in \cite{Blum:2013} with slight changes:
auxiliary likelihood is added and ``regularisation'' is placed under subset selection.
In practice the categories overlap, with some methods applying a combination of the strategies.

Subset selection and projection methods require a preliminary step of choosing a set of \emph{data features} $z(y)$.
For subset selection these can be thought of as \emph{candidate summary statistics}.
For convenience $z(y)$ is often written simply as $z$ below, and is a vector of scalar transformations of $y$, $(z_1, z_2, \ldots, z_k)$.
The \emph{feature selection} step of choosing $z$ is discussed further below.
Both methods also require \emph{training data} $(\theta_i, y_i)_{1 \leq i \leq n_{\text{train}}}$ to be created by simulation.

Subset selection methods select a subset of $z$, typically that which optimises some criterion on the training data.
Projection methods instead use the training data to choose a projection of $z$, for example a linear transformation $Az+b$, which performs dimension reduction.

Auxiliary likelihood methods take a different approach which does not need data features or training data.
Instead they make use of an approximating model whose likelihood (the ``auxiliary'' likelihood) is more tractable than the model of interest.
This may be chosen from subject area knowledge or make use of a general approach such as composite likelihood.
Summary statistics are derived from this approximating model, for example its the maximum likelihood estimators.

All these methods rely on some subjective input from the user.
In subset selection methods candidate summaries $z(y)$ must be supplied.
A typical choice will be a reasonably small set of summaries which are believed to be informative based on subject area knowledge.
(Large sets become too expensive for most methods to work with.)
Projection methods also require a subjective choice of $z(y)$.
A typical choice will be many interesting features of the data, and various non-linear transformations.
These may not be believed to be informative individually, but permit a wide class of potential projections.
There is less requirement for $\dim z$ to be small than for subset selection.
Auxiliary likelihood methods instead require the subjective choice of an approximate likelihood (discussed in Section \ref{sec:auxlik discussion}.)



\section{Subset selection methods} \label{sec:subsetselection}

Subset selection methods start with candidate summary statistics $z=(z_1, z_2, \ldots, z_k)$ and attempt to select an informative subset.
The methods below fall into two groups.
The first three run ABC for many possible subsets and choose the best based on information theoretic or other summaries of the output.
This requires ABC to be run many times, which is only computationally feasible for rejection or importance sampling ABC algorithms since these allow simulated datasets to be reused.
The final method, regularisation, takes a different approach.
All these methods are described in Section \ref{sec:ssmethods}.
Section \ref{sec:ssdiscussion} compares the methods and discusses the strengths and weaknesses of this strategy.

\subsection{Methods} \label{sec:ssmethods}

\paragraph{Approximate sufficiency \citep{Joyce:2008}}

\cite{Joyce:2008} propose a step-wise selection approach.
They add/remove candidate summary statistics to/from a subset one at a time and assess whether this significantly affects the resulting ABC posterior.
The motivation is that given sufficient statistics $S(\cdot)$ of minimal dimension,
adding further summaries will not change $\pi(\theta|S(\yobs))$ but removing any summary will.
This would be a test for sufficiency but requires perfect knowledge of $\pi(\theta|S(\yobs))$.
\citeauthor{Joyce:2008} propose a version of this test based on having only a density estimate and argue it is a test of \emph{approximate sufficiency}.
A further approximation is due to using $\pi_\ABC(\theta|S(\yobs))$ in place of $\pi(\theta|S(\yobs))$.

The approach involves testing whether the change from using summaries $S(y)$ to $S'(y)$ has a significant effect on the ABC posterior.
As various subsets are compared this test will be repeated under many choices of $S(y)$ and $S'(y)$.
A change is deemed significant if
\begin{equation}\label{eq:Rk}
\Bigg{|} \frac{\hat{\pi}_\ABC(\theta | S'(\yobs))}{\hat{\pi}_\ABC(\theta| S(\yobs))} -1 \Bigg{|} > T(\theta),
\end{equation}
where $\hat{\pi}_\ABC(\theta)$ is an estimated posterior density based on ABC rejection sampling output, detailed shortly.
The threshold $T(\cdot)$ is defined to test the null hypothesis that ABC targets the same distribution under both sets of summary statistics and control for multiple comparison issues arising from testing \eqref{eq:Rk} for several $\theta$ values.
See the appendix of \citeauthor{Joyce:2008} for precise details of how this is defined.

Only the case of scalar $\theta$ is considered by \citeauthor{Joyce:2008}.
Here they propose letting $\hat{\pi}_\ABC(\theta)$ be a histogram estimator.
That is, the support of $\theta$ is split into bins $B_1,B_2,\ldots,B_b$ and the proportion of the accepted sample in bin $B_i$ gives $\hat{\pi}_\ABC(\theta)$ for $\theta \in B_i$.
In practice \eqref{eq:Rk} is evaluated at a finite set of parameters $\theta_1 \in B_1, \theta_2 \in B_2, \ldots, \theta_b \in B_b$.

\citeauthor{Joyce:2008} note it is not obvious how to implement their method for higher dimensional parameters.
This is because the sample size required to use the histogram estimator of $\hat{\pi}_\ABC(\theta)$ becomes infeasibly large and the paper's choice of $T(\cdot)$ is specific to this estimator.

\paragraph{Entropy/loss minimisation \citep{Nunes:2010}}

\cite{Nunes:2010} propose two approaches.
The first aims to find the subset of $z$ which minimises the entropy of the resulting ABC posterior.
The motivation is that entropy measures how concentrated, and thus informative, the posterior is, with lowest entropy being most informative.
In practice an estimate of the entropy is used which is based on a finite sample from the ABC posterior: an extension of the estimate of \cite{Singh:2003}.

One criticism of the entropy criterion \citep{Blum:2013} is that in some circumstances an ABC posterior having smaller entropy does not correspond to more accurate inference.
For example it is possible that given a particularly precise prior the correct posterior may be more diffuse.
(See the next page for a further comment on this.)

The second approach aims to find the subset of $z$ which optimises the performance of ABC on datasets similar to $\yobs$, by minimising the average of the following loss function (root mean squared error)
\[
\left[ t^{-1} \sum_{i=1}^t|| \theta_i - \theta' ||_2 \right]^{1/2}.
\]
Here $\theta'$ is the parameter value which generated data $y'$,
and $(\theta_{i})_{1 \leq i \leq t}$ is the ABC output sample when $y'$ is used as the observations.
Performing this method requires generating $(\theta', y')$ pairs such that $y'$ is close to $\yobs$.
\citeauthor{Nunes:2010} recommend doing so using a separate ABC analysis whose summary statistics are chosen by entropy minimisation.
To summarise, this is a two-stage approach.
First select summaries by entropy minimisation and perform ABC to generate $(\theta',y')$ pairs.
Secondly select summaries by minimising root mean squared error.

\paragraph{Mutual information \citep{Barnes:2012, Filippi:2012}}

\cite{Barnes:2012} discuss how sufficiency can be restated in terms of the concept of \emph{mutual information}.
In particular, sufficient statistics maximise the mutual information between $S(y)$ and $\theta$.
From this they derive a necessary condition for sufficiency of $S(y)$:
the Kullback-Leibler (KL) divergence of $\pi(\theta | S(\yobs))$ from $\pi(\theta | \yobs)$ is zero i.e.
\[
\int \pi(\theta | \yobs) \log \frac{\pi(\theta | \yobs)}{\pi(\theta | S(\yobs))} d\theta = 0.
\]

This motivates a stepwise selection method to choose a subset of $z$.
A statistic $z_i$ is added to the existing subset $S(y)$ to create a new subset $S'(y)$
if the estimated KL divergence of $\pi_\ABC(\theta | S(\yobs))$ from $\pi_\ABC(\theta | S'(\yobs))$ is above a threshold.
One proposed algorithm chooses $z_i$ to maximise this divergence (a ``greedy'' approach) .
Another attempts to save time by selecting any acceptable $z_i$.
Steps are also provided to remove statistics that have become unnecessary.
Two approaches are given for estimating KL divergence between two ABC output samples.
The stepwise selection algorithm terminates when the improvement in KL divergence is below a threshold.
To determine a suitable threshold it is suggested to perform ABC several times with fixed summaries and evaluate the KL divergences between samples.
From this a threshold can be found indicating an insignificant change.

The mutual information method is closely related to the previous methods in this section.
Like the method of \cite{Joyce:2008} it seeks a subset $S(y)$ such that adding further statistics does not significantly change the ABC posterior.
However the KL criterion has the advantage that it can be applied when $\dim \theta > 1$.
It does share the disadvantage that $\pi_\ABC(\theta|S(\yobs))$ is used in place of $\pi(\theta|S(\yobs))$ which is a poor estimate unless $h \approx 0$.
Maximising mutual information as in \citeauthor{Barnes:2012} can be shown to be equivalent to minimising the \emph{expected} entropy of $\pi(\theta|S(y))$ (taking expectation over $y$).
This provides some information theoretic support for the entropy minimisation approach of \cite{Nunes:2010} but also gives more insight into its limitations: the entropy of $\pi(\theta|S(\yobs))$ which they consider may not be representative of expected entropy.
\citeauthor{Barnes:2012} also argue their method is more robust than \citeauthor{Nunes:2010}'s to the danger of selecting more statistics than is necessary for sufficiency.

\citeauthor{Barnes:2012} extend their method to model choice.
This is discussed in Section \ref{sec:modelchoice}.

\paragraph{Regularisation approaches \citep{Blum:2010b, Sedki:2012, Blum:2013}}

This method was proposed by \cite{Sedki:2012} and \cite{Blum:2013}.
The idea is to fit a linear regression with response $\theta$ and covariates $z$ based on training data and perform variable selection to find an informative subset of $z$.
Variable selection can be performed by minimising AIC or BIC
(see \citealp{Blum:2013} for details of calculated these in this setting.)
This typically requires a stepwise selection approach, although the cost of this could be avoided by using the lasso \citep{Hastie:2009} rather than ordinary regression.
The papers propose also using the fitted regression for ABC regression post-processing as well as summary statistic selection.
A related earlier approach \citep{Blum:2010b} proposed using a local linear regression model and performing variable selection by an empirical Bayes approach: maximising the likelihood of the training data after integrating out the distribution of the regression parameters.


\paragraph{Related methods}

\cite{Heggland:2004} provide asymptotic theory on summary statistic selection in another likelihood-free method, indirect inference.
Roughly speaking, the most useful summary statistics are those with low variance and expectation that changes rapidly with the parameters.
This theory is used to select summary statistics from a set of candidates by numerically estimating their variance and the derivative of their expectation based on a large number of simulations from promising parameter values.
It would be useful to develop ABC versions of this theory and methodology.
\cite{Ratmann:2007} go some way towards providing the latter for a particular application.

\subsection{Discussion} \label{sec:ssdiscussion}

\paragraph{Comparison of subset selection methods}

\cite{Joyce:2008} is based on rough ideas of sufficiency and can be implemented only in the limited setting of a scalar parameter.
The entropy minimisation method of \cite{Nunes:2010} and the approach of \cite{Barnes:2012} are successively more sophisticated information theoretic approaches.
The latter has the best theoretical support of methods based on such arguments.
However all these methods are motivated by properties of $\pi(\theta|S(\yobs))$ but then use $\pi_\ABC(\theta|S(\yobs))$ in its place.
For sufficiently large $h$ these may be quite different and it is unclear what effect this will have on the results.
The loss minimisation method of \cite{Nunes:2010} avoids this problem.
It chooses $S$ so that ABC results on simulated data optimise a specified loss function given a particular choice of $h$.
The question of robustness to the choice of which simulated data sets are used to assess this is still somewhat open, but otherwise this method seems to provide a gold standard approach.
The drawback of all the above methods is that they can be extremely expensive (see below).
Regularisation methods are cheaper but have received little study so their properties are not well understood.

\paragraph{Advantages and disadvantages of subset selection}

An advantage of subset selection methods is interpretability.
If the summaries in $z$ have intuitive interpretations, then so will the resulting subset.
This is especially useful for model criticism by the method of \cite{Ratmann:2009}.
Here, one investigates whether the simulated $S(y)$ values accepted by ABC are consistent with $\sobs$.
If some particular component of $\sobs$ cannot be matched well by the simulations, this suggests model misspecification.
Understanding the interpretation of this summary can hopefully suggest model improvements.

A disadvantage is the implicit assumption that a low dimensional informative subset of $z$ exists.
Subset selection can be thought of as a projection method which is restricted to projections to subsets.
However, it may be the case that the best choice is outside this restriction
e.g.~the mean of the candidate summaries.

Further disadvantages are cost and scalability problems.
For small $k$ (denoting $\dim z$) it may be possible to evaluate the performance of all subsets of $z$ and find the global optimum.
For large $k$ this is prohibitively expensive and more sophisticated search procedures such as stepwise selection must be used.
However, such methods may only converge to a local optimum
and the computational cost still grows rapidly with $k$.

Computational cost is particularly large for the methods which require ABC to be rerun for each subset of $z$ considered.
The computing requirement is typically made more manageable by using the same simulations for each ABC analysis.
However this restricts the algorithm used in this stage to ABC rejection or importance sampling.
Therefore the resulting summary statistics have not been tested at the lower values of $h$ which could be achieved using ABC-MCMC or ABC-SMC, and may not be good choices under these algorithms.

Finally, as discussed in Section \ref{sec:strategies}, subset selection methods require a feature selection stage to choose the set of potential summaries $z$.
In all the papers cited above this step is based on subjective choice and is crucial to good performance.
Comparable subjective choices are also required by the strategies described later.
However a particular constraint here is that $\dim z$ cannot be too large or the cost of subset selection becomes infeasible.

\section{Projection methods} \label{sec:projection}

Projection methods start with a vector of data features $z(y)=(z_1,z_2,\ldots,z_k)$ and attempts to find an informative lower dimensional projection, often a linear transformation.
To choose a projection, training data $(\theta_i, y_i)_{1 \leq i \leq n_{\text{train}}}$ is created by simulation from the prior and model and some dimension reduction technique is applied.
Section \ref{sec:dimred} presents various dimension reduction methods which have been proposed for use in ABC.
Section \ref{sec:impldetails} describes variations in how the training data is generated.
It is generally possible to match up the two parts of the methodology as desired.
Section \ref{sec:projdiscussion} compares the methods and discusses the strengths and weaknesses of this strategy.

\subsection{Dimension reduction techniques} \label{sec:dimred}

\paragraph{Partial least squares \citep{Wegmann:2009}}

Partial least squares (PLS) aims to produce linear combinations of covariates which have high covariance with responses and are uncorrelated with each other.
In the ABC setting the covariates are $z_1,\ldots,z_k$ and the responses are $\theta_1,\ldots,\theta_p$.
The $i$th PLS component $u_i = \alpha_i^T z$ maximises
\[
\sum_{j=1}^p \Cov(u_i, \theta_j)^2,
\]
subject to $\Cov(u_i,u_j)=0$ for $j < i$ and a normalisation constraint on $\alpha_i$ such as $\alpha_i^T \alpha_i=1$.
In practice empirical covariances based on training data are used.
Several algorithms to compute PLS components are available.
These can produce different results as they use slightly different normalisation constraints.
For an overview of PLS see \cite{Boulesteix:2007}.

PLS produces $\min(k,n_{\text{train}}-1)$ components.
\cite{Wegmann:2009} use the first $c$ components as ABC summary statistics, with $c$ chosen by a cross-validation procedure.
This aims to minimise the root mean squared error in a linear regression of $\theta$ on $u_1, \ldots, u_c$.
This approach is similar to the regularisation subset selection methods of Section \ref{sec:subsetselection}.

\paragraph{Linear regression \citep{Fearnhead:2012}}

\cite{Fearnhead:2012} fit a linear model to the training data: $\theta \sim N(A z + b, \Sigma)$.
The resulting vector of parameter estimators $\hat{\theta}(y)=A z + b$ is used as ABC summary statistics.
This is a low dimensional choice: $\dim \hat{\theta}(y)  = \dim \theta = p$.

Motivation for this approach comes from the following result.
Consider $\pi(\theta | S(\yobs))$, which is the ABC target distribution for $h=0$.
Then $S(y)=\E(\theta|y)$ can be shown to be the optimal choice in terms of minimising quadratic loss of the parameter means in this target distribution.
Fitting a linear model produces an estimator of these ideal statistics.

A linear regression estimate of $\E(\theta|y)$ is crude, but can be improved by selecting good $z(y)$ features.
\citeauthor{Fearnhead:2012} propose comparing $z(y)$ choices by looking at the goodness of fit of the linear model, in particular the BIC values.
Another way to improve the estimator is to train it on a local region of the parameter space.
This is discussed in Section \ref{sec:impldetails}.
\citeauthor{Fearnhead:2012} use the name ``semi-automatic ABC'' to refer to summary statistic selection by linear regression with these improvements.
Good performance is shown for a range of examples, and is particularly notable when $z(y)$ is high dimensional \citep{Fearnhead:2012, Blum:2013}.

As \cite{Robert:2012} points out, the theoretical support for this method is only heuristic as it focuses on the unrealistic case of $h=0$.
Another limitation is that these summaries focus on producing good point estimates of $\theta$, and not on accurate uncertainty quantification.
\citeauthor{Fearnhead:2012} propose a modified ABC algorithm (``noisy ABC'') which tackles this problem to some extent.

The discussion above also motivates using more advanced regression-like methods.
\citeauthor{Fearnhead:2012} investigate the lasso \citep{Hastie:2009}, canonical correlation analysis \citep{Mardia:1979} and sliced inverse regression \citep{Li:1991} (see \citealp{Prangle:2011} for details).
The former two do not produce significant improvements over linear regression in the applications considered.
The latter produces large improvements in one particular example, but requires significant manual tuning.
Many further suggestions can be found in the discussions published alongside \cite{Fearnhead:2012}.

\paragraph{Boosting \citep{Aeschbacher:2012}}

Boosting is a non-linear regression method.
Like linear regression it required training data and outputs predictors $\hat{\theta}(y)$ of $\E(\theta|y)$ which can be used as ABC summary statistics.
Boosting is now sketched for scalar $\theta$.
For multivariate $\theta$ the whole procedure can be repeated for each component.
The approach begins by fitting a ``weak'' estimator to the training data.
For this \cite{Aeschbacher:2012} use linear regression with response $\theta$ and a single covariate: whichever feature in $z(y)$ maximises reduction in error (e.g.~mean squared error).
The training data is then weighted according to its error under this estimator.
A second weak estimator is fitted to this weighted training data and
a weighted average of the first two estimators is formed.
The training data is weighted according to its error under this, and a third weak estimator is formed, and so on.
Eventually the process is terminated and the final weighted average of weak estimators is output as a ``strong'' estimator.
The idea is that each weak estimator attempts to concentrate on data which has been estimated poorly in previous steps.
See \cite{Buhlmann:2007} for a full description.

\subsection{Generating training data} \label{sec:impldetails}

A straightforward approach to draw training data pairs $(\theta, y)$ is to sample $\theta$ from the prior and $y$ from the model conditional on $\theta$.
This approach is used by \cite{Wegmann:2009} for example.
In rejection or importance sampling ABC algorithms this training data can be reused to implement the ABC analysis.
Hence there is no computational overhead in producing training data.
For other ABC algorithms this is not the case.

\cite{Fearnhead:2012} and \cite{Aeschbacher:2012} use different approaches to generate training data which aim to make the projection methods more effective.
The idea is that the global relationship between $\theta$ and $y$ is likely to be highly complicated and hard to learn.
Learning about the relationship close to $\yobs$ may be easier.
This motivates sampling training pairs from a more concentrated distribution.

\cite{Aeschbacher:2012} implement this by performing a pilot ABC analysis using $S(y)=z$ (i.e.~all the data features).
The accepted simulations are used as training data for their boosting procedure.
The resulting summary statistics are then used in an ABC analysis.

\cite{Fearnhead:2012} argue that such an approach might be dangerous.
This is because $S(y)$ is only trained to perform well on a concentrated region of $(\theta,y)$ values, and could perform poorly outside this region.
In particular it is possible that $S(y) \approx S(\yobs)$ in regions excluded by the pilot analysis, producing artefact posterior modes.
\citeauthor{Fearnhead:2012} instead recommend performing a pilot ABC analysis using ad-hoc summary statistics.
This is used to find a \emph{training region} of parameter space, $R$, containing most of the posterior mass.
Training $\theta$ values are drawn from the prior truncated to $R$, and $y$ values from the model.
Summary statistics are fitted and used in a main ABC analysis which also truncates the prior to $R$.
This ensures that $\theta$ regions excluded by the pilot remain excluded in the main analysis.
Note that this truncation approach was introduced by \cite{BlumFrancois:2010} in a regression post-processing context.

\subsection{Discussion} \label{sec:projdiscussion}

\paragraph{Comparison of projection methods}

Partial least squares is a well established dimensional reduction method.
However it does not have any theoretical support for use in ABC and sometimes performs poorly in practice \citep{Blum:2013}.
\cite{Fearnhead:2012} provide heuristic theoretical support to the approach of constructing parameter estimators for use as ABC summary statistics, and show empirically implementing this by the simple method of linear regression can perform well in practice.
It is likely that more sophisticated regression approaches will perform even better.
Boosting is one example of this.
A particularly desirable goal would a regression method which can incorporate the feature selection step and estimate $E(\theta|y)$ directly from the raw data $y$.
This is discussed further in Section \ref{sec:discussion}.

\paragraph{Advantages and disadvantages of projection methods}

Projection methods avoid some of the disadvantages of subset selection methods.
In particular the high computational costs associated with repeating calculations for many possible subsets are avoided.
Also a wider space of potential summaries is searched, not just subsets of $z$.
However this means that the results may be less interpretable and thus harder to use for model criticism.
(For further discussion of all these points see Section \ref{sec:ssdiscussion}.)

Another advantage of projection methods is that they can be implemented on almost any problem.
This is in contrast to auxiliary likelihood methods which require the specification of a reasonable approximate likelihood.

Projection methods require a subjective choice of features $z(y)$, as do subset selection methods.
However projection methods have more freedom to choose a large set of features and still have a feasible computational cost, and some methods provide heuristic tools to select between feature sets (i.e.~\citealp{Fearnhead:2012} use BIC.)

\section{Auxiliary likelihood methods} \label{sec:auxlik}

An intuitively appealing approach to choosing summary statistics for ABC inference of a complicated model is to use statistics which are known to be informative for a simpler related model.
This has been done since the earliest precursors to ABC in population genetics \citep[e.g.][]{Fu:1997, Pritchard:1999}.
Recently there has been much interest in formalising this approach.
The idea is to specify an approximate and tractable likelihood for the data, referred to as an \emph{auxiliary likelihood}, and derive summary statistics from this.
Several methods to do this have been proposed which are summarised in Section \ref{sec:auxlik methods} and discussed in Section \ref{sec:auxlik discussion}.
There have also been related proposals for new likelihood-free methods based on auxiliary likelihoods, which are covered elsewhere in this volume (see chapters 8 and 12).

First some notation and terminology is introduced.
The auxiliary likelihood is represented as $p_A(y|\phi)$.
This can be thought of as defining an \emph{auxiliary model} for $y$.
This differs from the model whose posterior is sought, which is referred to here as the \emph{generative model}.
The auxiliary model parameters, $\phi$, need not correspond to those of the generative model, $\theta$.
A general question is which auxiliary likelihood to use.
This is discussed in Section \ref{sec:auxlik discussion}, including a description of some possible choices.
To assist in reading Section \ref{sec:auxlik methods} it may be worth keeping in mind the simplest choice: let the auxiliary model be a tractable simplified version of the generative model.

\subsection{Methods} \label{sec:auxlik methods}

\paragraph{Maximum likelihood estimators (ABC-IP)}

Here $S(y) = \hat{\phi}(y) = \argmax_\phi p_A(y|\phi)$.
That is, the summary statistic vector is the maximum likelihood estimator (MLE) of $\phi$ given data $y$ under the auxiliary model.
To use this method this MLE must be unique for any $y$.
Typically $S(y)$ must be calculated numerically, which is sometimes computationally costly.

This approach was proposed by \cite{Drovandi:2011}, although \cite{Wilson:2009} use a similar approach in a particular application.
It was motivated by a similar choice of summaries in another likelihood-free method, \emph{indirect inference} \citep{Gourieroux:1993}.
The terminology ABC-IP for this approach was introduced by \cite{Gleim:2013}: ``I'' represents indirect and ``P'' using parameter estimators as summaries.


Some theoretical support for ABC-IP is available.
\cite{Gleim:2013} note that classical statistical theory shows that $\hat{\phi}(y)$ is typically \emph{asymptotically sufficient} for the auxiliary model (see chapter 9 of \citealp{Cox:1979} for full details).
This assumes an asymptotic setting where $n \to \infty$ as the data becomes more informative.
As a simple example $y$ could consist on $n$ independent identically distributed observations.
Asymptotic sufficiency implies $\hat{\phi}(y)$ asymptotically summarises all the information about the auxiliary model parameters.

Ideally $\hat{\phi}(y)$ would also be asymptotically sufficient for the generative model.
\citeauthor{Gleim:2013} show this is the case if the generative model is nested within the auxiliary model.
However having a tractable auxiliary model of this form is rare.
\cite{Martin:2014} note that even without asymptotic sufficiency for the generative model, \emph{Bayesian consistency} can be attained.
That is, the distribution $\pi(\theta|\hat{\phi}(y))$ asymptotically shrinks to a point mass on the true parameter value.
They give necessary conditions for consistency: essentially that in this limit $\hat{\phi}(y)$ perfectly discriminates between data generated by different values of $\theta$.

\paragraph{Likelihood distance (ABC-IL)}

\cite{Gleim:2013} suggest a variation on ABC-IP which uses the distance function:
\begin{equation} \label{eq:ABCILdist}
|| \hat{\phi}(y), \hat{\phi}(\yobs) || = \log p_A(\yobs | \hat{\phi}(\yobs)) - \log p_A(\yobs | \hat{\phi}(y)).
\end{equation}
This is the log likelihood ratio for the auxiliary model between the MLEs under the observed and simulated datasets.
They refer to this as ABC-IL: ``L'' represents using a likelihood distance.

It is desirable that $|| \hat{\phi}(y), \hat{\phi}(\yobs) || = 0$ if and only if $\hat{\phi}(y) = \hat{\phi}(\yobs)$.
This requires $p_A$ to be well behaved.
For example it suffices that $y \mapsto \hat{\phi}(y)$ is a one-to-one mapping.
However weaker conditions can sometimes be used:
see \cite{Drovandi:2015} section 7.3 for example.

\paragraph{Scores (ABC-IS)}

\cite{Gleim:2013} suggest taking
\begin{equation} \label{eq:score}
S(y) = \left( \tfrac{\partial}{\partial \phi_i} \log p_A(y|\phi) \middle|_{\phi=\hat{\phi}(\yobs)} \right)_{1 \leq i \leq p}.
\end{equation}
This is the score of the auxiliary likelihood evaluated under parameters $\hat{\phi}(\yobs)$.
As earlier $\hat{\phi}(\yobs)$ is the MLE of $\yobs$ under the auxiliary likelihood.
\citeauthor{Gleim:2013} refer to this approach as ABC-IS: ``S'' refers to using score summaries.

The motivation is that the score has similar asymptotic properties to those described above for the MLE \citep{Gleim:2013, Martin:2014} but is cheaper to calculate.
This is because numerical optimisation is required once only, to find $\hat{\phi}(\yobs)$, rather than every time $S(y)$ is computed.
\cite{Drovandi:2015} also note that ABC-IS is more widely applicable than ABC-IP as it does not require existence of a unique MLE for $p_A(y|\phi)$ under all $y$, only under $\yobs$.



Some recent variations of ABC-IS include:
application to state-space models by using a variation on Kalman filtering to provide the auxiliary likelihood, and use of a \emph{marginal score} \citep{Martin:2014};
alternatives to the score function \eqref{eq:score} based on \emph{proper scoring rules} \citep{Ruli:2014};
and using a \emph{rescaled score} when the auxiliary model is a composite likelihood \citep{Ruli:2015}.


\subsection{Discussion} \label{sec:auxlik discussion}

\paragraph{Comparison of auxiliary likelihood methods}

ABC-IS has the advantage that it is not based on calculating the MLE repeatedly.
This can be computationally costly, may be prone to numerical errors, and indeed a unique MLE may not even exist.
Furthermore, the asymptotic properties discussed above suggest the score-based summaries used by ABC-IS encapsulate similar information about the auxiliary likelihood as the MLE.
This recommends use of ABC-IS.
However empirical comparisons by \cite{Drovandi:2015} suggest the best auxiliary likelihood method in practice is problem specific (see Section \ref{sec:discussion} for more details).

\paragraph{Which auxiliary likelihood?}

Various choices of auxiliary likelihood have been used in the literature.
Examples include the likelihood of a tractable alternative model for the data, or of a flexible general model such as a Gaussian mixture.
Another is to use a tractable approximation to the likelihood of the generative model such as composite likelihood \citep{Varin:2011}.
There is a need to decide which choice to use.

An auxiliary likelihood should ideally have several properties.
To produce low dimensional summary statistics, it should have a reasonably small number of parameters.
Also it is desirable that it permits fast and accurate computation of the MLE or score.
These two requirements are easy to assess.
A third requirement is that the auxiliary likelihood should produce summary statistics which are informative about the generative model.
This seems harder to quantify.

\cite{Drovandi:2015} recommend performing various goodness-of-fit tests to see how well the auxiliary likelihood matches the data.
Similarly \cite{Gleim:2013} use the BIC to choose between several possible auxiliary likelihoods.
Such tests are computationally cheap and give insight into the ability of the model to summarise important features of the data.
However it is not clear that performing better in a goodness-of-fit test necessarily results in a better ABC approximation.
Ideally what is needed is a test of how well an auxiliary likelihood discriminates between datasets drawn from the generative model under different parameter values.
How to test this is an open problem.

\paragraph{Advantages and disadvantages of auxiliary likelihood methods}

Auxiliary likelihood methods avoid the necessity of choosing informative data features required by subset selection and projection methods.
This is replaced by the somewhat analogous need to choose an informative auxiliary likelihood.
However such a choice may often be substantially easier, particularly if well-developed tractable approximations to the generative model are already available.
In others situations, both tasks may be equally challenging.

Another advantage of auxiliary likelihood methods is that they avoid the computational cost of generating of training data, as required by preceding methods.
Instead they make use of subject area knowledge to propose auxiliary likelihoods.
In the absence of such knowledge one could try to construct an auxiliary likelihood from training data.
This is one viewpoint of how projection methods based on regression operate.

\section{Model choice} \label{sec:modelchoice}

ABC can be applied to inference when there are several available models $M_1, M_2, \ldots, M_r$.
See Chapter 7 in this volume or \cite{Didelot:2011} for details of algorithms.
This section is on the problem of choosing which summary statistics to use here.
The aim of most work to date is to choose summaries suitable for inferring the posterior model weights.
The more challenging problem of also inferring model parameters is mentioned only briefly.

A natural approach used by some early practical work is to use summary statistics which are informative for parameter inference in each model.
Unfortunately, except in a few special cases, this can give extremely poor model choice results as highlighted by \cite{Robert:2011}.
The issue is that summary statistics which are good for parameter inference within models are not necessarily informative for choosing between models.
This section summarises more recent theoretical and practical work which shows that informative summary statistics can be found for ABC model choice.
Therefore ABC model choice can now be trusted to a similar degree to ABC parameter inference.

The remainder of this section is organised as follows.
Section \ref{sec:mc theory} re-examines sufficiency and other theoretical issues for ABC model choice, as there are some surprisingly different results to those described earlier in the chapter for the parameter inference case.
Section \ref{sec:mc methods} reviews practical methods of summary statistic selection
and Section \ref{sec:mc discussion} gives a brief discussion.

Note that model choice can be viewed as inference of a discrete parameter $m \in \{1,2,\ldots,r\}$ indexing the available models.
Therefore the following material would also apply to ABC inference of a discrete parameter.
However, as elsewhere in this chapter, the phrase ``parameter inference'' is generally used in the section as shorthand to refer to inference of \emph{continuous} parameters.

\subsection{Theory} \label{sec:mc theory}

\paragraph{Curse of dimensionality}

As described earlier ABC suffers a curse of dimensionality when dealing with high dimensional data.
Theoretical work on this, summarised in Section \ref{sec:cod} has focused on the parameter inference case.
However the technical arguments involved focus on properties of the summary statistics, rather than of the parameters.
Therefore it seems likely that the arguments can be adapted to give unchanged results for model choice simply by considering the case of discrete parameters.

This means it remains important to use low dimensional summary statistics for ABC model choice.

\paragraph{Sufficiency and consistency}

As for parameter inference case the ideal summaries for ABC model choice would be low dimensional sufficient statistics.
Unlike the case of parameter inference such statistics do exist for ABC model choice, and results are also available on links to consistency and sufficiency for parameter inference.
These theoretical results are now summarised, and will motivate some of the methods for summary statistic choice described in the next section.

First some terminology is defined, based on the definitions of sufficiency in Section \ref{sec:sufficiency}.
Let $\theta_i$ represent the parameters associated with model $M_i$.
Statistics $S(y)$ that are sufficient for $\theta_i$ under model $M_i$ will be referred to below as \emph{sufficient for parameter inference} in that model.
Now consider the problem of jointly inferring $\theta_1, \theta_2, \ldots, \theta_r, m$, where $m$ is a model index.
This is equivalent to inference on an encompassing model in which the data is generated from $M_i$ conditional on $\theta_i$ when $m=i$.
Statistics will be referred to as \emph{sufficient for model choice} if they are Bayes sufficient for $m$ in this encompassing model.

\cite{Didelot:2011} show that sufficient statistics for model choice between models $M_1$ and $M_2$ can be found by taking parameter inference sufficient statistics of a model in which both are nested.
This result is of limited use as such parameter inference sufficient statistics rarely exist in low dimensional form (see discussion in Section \ref{sec:sufficiency}).
However it has useful consequences in the special case where $M_1$ and $M_2$ are both exponential family distributions i.e.~
\[
\pi(y|\theta_i,M_i) \propto \exp\left[ s_i(x)^T \theta_i + t_i(x) \right]
\]
for $i=1,2$.
In this case $s_i(x)$ is a vector of parameter inference sufficient statistics for model $M_i$ and $\exp[t_i(x)]$ is known as the base measure.
\citeauthor{Didelot:2011} show that sufficient statistics for model choice are the concatenation of $s_1(x), s_2(x), t_1(x)$ and $t_2(x)$.

\cite{Prangle:2014} prove that the following vector of statistics is sufficient for model choice
\begin{align*} 
T(y)&=(T_1(y),T_2(y),\ldots,T_{r-1}(y)), \\
\text{where} \quad T_i(y)&=\pi(y|M_i)/\sum_{j=1}^r \pi(y|M_j).
\end{align*}
Here $T_i(y)$ is the evidence under $M_i$ divided by the sum of model evidences.
Furthermore any other vector of statistics is sufficient for model choice if and only if it can be transformed to $T(y)$.

Thus low dimensional sufficient statistics exist if the number of models $r$ is reasonably small.
This results may seem at first to contradict the arguments of Section \ref{sec:theory} that these are only available for exponential family models.
A contradiction is avoided because model choice is equivalent to inferring the discrete parameter $m$ and a model with a discrete parameter can be expressed as an exponential family.

A related result is proved by \cite{Marin:2014}.
They give necessary conditions on summary statistics $S(y)$ for $\Pr(m|S(y))$ to be consistent in an asymptotic regime corresponding to highly informative data.
That is, these conditions allow for perfect discrimination between models in this limiting case.
In addition to several technical conditions,
the essential requirement is that the limiting expected value of the summary statistic vector should differ under each model.

\subsection{Methods} \label{sec:mc methods}

\paragraph{Using an encompassing model \citep{Didelot:2011}}

As described above, \cite{Didelot:2011} prove that sufficient statistics for model choice between exponential family models can be found by concatenating the parameter sufficient statistics and base measures of each model.
Situations where this can be used are rare, but one is the Gibbs random field application considered by \cite{Grelaud:2009}.
In this case the base measures are constants and so can be ignored.

\paragraph{Mutual information \citep{Barnes:2012}}

This method was described earlier for the case of parameter inference.
To recap briefly, it is a subset selection method motivated by the concept of mutual information which sequentially adds or removes candidate summary statistics to or from a set.
Each time a Kullback-Leibler divergence between the ABC posterior distributions under the previous and new sets is estimated.
The process terminates when the largest achievable divergence falls below a threshold.

\cite{Barnes:2012} adapt this method to find summary statistics for joint model and parameter inference as follows.
First they estimate sufficient statistics for parameter inference under each model, and concatenate these.
Next they add further statistics until model sufficiency is also achieved.
Alternatively the method could easily be adapted to search for statistics which are sufficient for model choice only.

\paragraph{Projection/classification methods \citep{Estoup:2012, Prangle:2014}}

The idea here is to use training data to construct a classifier which attempts to discriminate between the models given data $y$.
Informative statistics are taken from the fitted classifier and used as summary statistics in ABC.
This can be thought of as a projection approach mapping high dimensional data $y$ to low dimensional summaries.

Two published approaches of this form are now described in more detail.
Training data $(\theta_i, m_i, y_i)_{1 \leq i \leq n_{\text{train}}}$ is created where $y_i$ is drawn from $\pi(y|\theta_i, m_i)$ (generating $\theta_i, m_i$ pairs is discussed below).
A vector of data features $z(y)$ must be specified.
A classification method is then used to find linear combinations $\alpha^T z(y)$ which are informative about $m$, the model index.
\cite{Estoup:2012} use linear discriminant analysis and \cite{Prangle:2014} use logistic regression (for the case of two models).
For a review of both see \cite{Hastie:2009}, who note they typically give very similar results.
As motivation \citeauthor{Prangle:2014} observe that, in the two model case, logistic regression produces a crude estimate of $\logit [\Pr(M_1|y)]$, which would be sufficient for model choice as discussed above (extending this argument to more than two models is also discussed).

The simplest approach to drawing $\theta_i,m_i$ pairs is simply to draw $m_i$ from its prior (or with equal weights if this is too unbalanced) and $\theta_i$ from the appropriate parameter prior.
\citeauthor{Prangle:2014} observe that it can sometimes be hard for their classifier to fit the resulting training data.
They propose instead producing training data which focuses on the most likely $\theta$ regions under each model (as in the similar approach for parameter inference in Section \ref{sec:impldetails}).
The resulting summary statistics are only trained to discriminate well in these regions, so a modified ABC algorithm is required to use them.
This involves estimation of some posterior quantities and so may not be possible in some applications.

Alternatively, \citeauthor{Estoup:2012} first perform ABC with a large number of summary statistics.
The accepted output is used as training data to fit model choice summary statistics.
These are then used in regression post-processing.
This avoids the need for a modified ABC algorithm, but the first stage of the analysis may still suffer from errors due to the curse of dimensionality.



\paragraph{Local error rates \citep{Stoehr:2014}}

\cite{Stoehr:2014} compare three sets of summary statistics for ABC model choice in the particular setting of Gibbs random fields.
The idea is to pick the choice which minimises the \emph{local error rate}, defined shortly.
This method could easily be used more generally, for example as the basis of a subset selection method similar to the loss minimisation method of \cite{Nunes:2010}.

The local error rate is $\Pr(\hat{M}(\yobs) \neq M | \yobs)$, where $\hat{M}(y)$ is the model with greatest weight under the target distribution of the ABC algorithm given data $y$.
(This can be interpretted as using a 0-1 loss function).
In practice this quantity is unavailable, but it can be estimated.
Suppose a large number of $(y_i, M_i)_{1 \leq i \leq n_{\text{val}}}$ \emph{validation} pairs have been generated.
\citeauthor{Stoehr:2014} suggest running ABC using each $y_i$ in turn as the observations and evaluating an indicator variable $\delta_i$ which equals 1 when ABC assigns most weight to model $M_i$.
Non-parametric regression is then used to estimate $\Pr(\delta=1 | \yobs)$.
This is challenging if $\dim y$ is large, so dimension reduction is employed.
\citeauthor{Stoehr:2014} use linear discriminant analysis for this (as in the \citealp{Estoup:2012} approach described above.)
To reduce costs, a cross-validation scheme is used so that the same simulations can be used for ABC analyses and validation.





\subsection{Discussion} \label{sec:mc discussion}

\paragraph{Comparison of methods}

The approach of \cite{Didelot:2011} -- choosing sufficient statistics for an encompassing model -- is only useful in specialised circumstances, such as choice between exponential family models.
The other methods listed above are more generally applicable.
Their advantages and disadvantages are similar to those discussed earlier for corresponding parameter inference methods.
In particular, the two subset selection methods have the disadvantage that they have high computational costs if there are many candidate summary statistics.

\paragraph{Prospects}

Comparatively few summary statistic selection methods have been proposed for the model choice setting.
Thus there is potential to adapt other existing approaches from the parameter inference case for use here.
In particular, it would be interesting to see whether regularisation or auxiliary likelihood approaches can be developed.

Another promising future direction is to construct model choice summary statistics using more sophisticated classification methods than those described above, for example random forests or deep neural networks.
As an alternative to using these methods to produce summary statistics, some of them can directly output likelihood-free inference results \citep{Pudlo:2014}.

Finally, choosing summary statistics for joint ABC inference of model and parameters is a desirable goal.
One approach is to separately find summaries for model choice and for parameter inference in each model and concatenate these.
However it may be possible to produce lower dimensional informative summaries by utilising summaries which are informative for several of these goals.
Finding methods to do this is an open problem.

\section{Discussion} \label{sec:discussion}

\paragraph{Empirical performance}

Most papers proposing methods of summary statistic choice report some empirical results on performance.
These show some merits to all the proposed methods.
However it is difficult to compare these results to each other as there are many differences between the applications, algorithms and tuning choices used.
Two studies are reported here which compare several methods on multiple parameter inference applications.
Little comparable work exists for model choice methods.

\cite{Blum:2013} compare several subset selection and projection methods on three applications using ABC rejection sampling.
They conclude: ``What is very apparent from this study is that there is no single `best' method of dimension reduction for ABC.''
The best performing methods for each application are:
the two stage method of \cite{Nunes:2010} on the smallest example ($k = 6$).
the AIC and BIC regularisation methods on a larger example ($k = 11$)
and the linear regression method of \cite{Fearnhead:2012} on the largest example ($k = 113$).
(Recall $k=\dim z$ i.e.~the number of data features.)

\cite{Drovandi:2015} compare auxiliary likelihood methods on several applications using ABC MCMC.
They conclude: ``Across applications considered in this paper, ABC IS was the most
computationally efficient and led to good posterior approximations.''
However they note that its posterior approximation was not always better than ABC-IP and ABC-IL, so that again the best approach seems problem specific.

\paragraph{Which method to use?}

Although many methods for choosing summary statistics have been proposed, there are no strong theoretical or empirical results about which to use in practice for a particular problem.
Also the area is developing rapidly, and many new approaches can be expected to appear soon.
Therefore only very general advice is given here.

Each of the strategies discussed has its particular strengths.
When a small set of potential summaries can be listed, subset selection performs a thorough search of possible subsets.
When a good tractable approximate likelihood is available, auxiliary likelihood methods can make use of it to produce informative parameter inference summaries, although they are not yet available for model choice.
Projection methods are highly flexible and can be applied to almost any problem.

It seems advisable to consider subset selection or auxiliary likelihood methods in situations that suit their strengths, and projection methods otherwise.
The question of which methods are most appealing within each strategy is discussed within their respective sections.
For parameter inference the empirical comparisons described above can also provide some guidance.

Ideally, if resources are available, the performance of different methods should be assessed on the problem of interest.
This requires repeating some or all of the analysis for many simulated datasets.
To reduce the computation required, \cite{Bertorelle:2010} and \cite{Blum:2013} advocate performing a large set of simulations and reusing them to perform the required ABC analyses.
This restricts the algorithm to ABC rejection or importance sampling.

Finally note that there is considerable scope to modify the summary statistics generated by the methods in this chapter.
For example the user may decide to choose a combination of statistics produced by several different methods,
or add further summary statistics based on subject area insights.

\paragraph{Prospects}

This chapter has shown how, amongst other approaches, classification and regression methods can be used to provide ABC summary statistics from training data.
There are many sophisticated tools for these in the statistics and machine learning literature which may produce more powerful ABC summary statistic selection methods in future.
It would be particularly desirable to find methods which do not require a preliminary subjective feature selection stage.
One promising approach to this is regression using deep neural networks \citep{Bengio:2015}, although it is unclear whether the amount of training data required to fit these well would be prohibitively expensive to simulate.
Another possibility is to come up with dictionaries of generally informative features for particular application areas.
\cite{Fulcher:2013} and \cite{Stocks:2014} implement ideas along these lines for time series analysis and population genetics respectively.

A topic of recent interest in ABC is the case where a dataset $y$ is made up of many observations
which are either independent and identically distributed, or have some weak dependence structure.
Several approaches to judging the distance between such datasets have been recently proposed.
These fall somewhat outside the framework of this chapter, as they bypass producing conventional summary statistics and instead simply define a distance.
An interesting question is the extent to which these alleviate the curse of dimensionality.
The methods base distances on:
classical statistical tests \citep{Ratmann:2013};
the output of classifiers \citep{Gutmann:2014};
kernel embeddings \citep{Park:2015}.

This chapter has concentrated on using summary statistics to reduce the ABC curse of dimensionality and approximate the true posterior $\pi(\theta|\yobs)$.
However other aspects of summary statistics are also worth investigating.
Firstly it is possible that dimension-preserving transformations of $S(y)$ may also improve ABC performance.
This is exploited by \cite{Yildirim:2014} in the context of a specific ABC algorithm for example.
Secondly, several authors \citep{Wood:2010,Girolami:2012,Fasiolo:2014} discuss cases where the true posterior is extremely hard to explore, for example because it is very rough with many local modes.
They argue that using appropriate summary statistics can produce a better behaved $\pi(\theta|\sobs)$ distribution which is still informative about model properties of interest.

\bibliography{sumstats}

\end{document}